# IMPLEMENTATION OF MICROWAVE WITH ARBITRARY AMPLITUDE AND PHASE FOR THE DCLS

H.K. Li[1], H.L. Ding[1†], Y. Li, J.F. Zhu[‡], J.W. Han, X.W. Dai, J.Y. Yang[1], W.Q. Zhang[1]
Institute of Advanced Science Facilities, Shenzhen, China
[1] also at Dalian Institute of Chemical Physics, Chinese Academy of Sciences, Dalian, China

## Abstract

In many experiments, the simultaneous emission of multiple wavelengths of FEL (Free-Electron Laser) is significant. For the pulsed-mode FEL facility, we must accelerate multiple electron beams in one microwave pulse, and they may be in different amplitudes and phases in the acceleration field. Therefore, we implement a microwave excitation, whose amplitude and phase have arbitrary shapes in the LLRF (Low-Level Radiofrequency) system. We generate a microwave pulse with step-shaped amplitude and phase for dual beam operation in DCLS (Dalian Coherent Light Source). The microwave system of the primary accelerator has four pulsed LLRF devices, which output excitation to drive four solid-state amplifiers and then excite two 50 MW and two 80 MW klystrons, respectively. Preliminary experiments have shown that this step-shaped microwave can be used for the DCLS twin-bunch operation.

## MOTIVATION

With the development of the FEL (Free-Electron Laser) application, over the years, two-colour FEL has been demanded by pump-probe experiments. This technique can be used to research molecular structure, planetary interior physics, fusion drive etc. In those experiments, the simultaneous emission of multiple wavelengths of FEL (Free-Electron Laser) is significant. For the pulsed-mode FEL facility, multiple electron beams must be accelerated in one microwave pulse in different amplitudes and phases in the acceleration field. Therefore, we implement a microwave excitation, whose amplitude and phase have arbitrary shapes in the LLRF (Low-Level Radiofrequency) system.

## METHOD AND ALGORITHM

The resonance relation of FEL is as follows:

$$\lambda = \frac{\lambda_u}{2\gamma^2}\left(1 + \frac{\kappa^2}{2}\right) \quad (1)$$

The $\lambda_u$ is the wavelength of undulator field, the $\gamma$ is the Lorentz factor of the beam, the $\kappa$ is the parameter of undulator, the $\lambda$ is the central wavelength of the laser pulse. Currently, there are two basic functions to realize the two-colour FEL: single beam two-colour FEL [1] and twin-bunch FEL [2]. Although the previous can well control the timing and spectral characteristics of laser pulses, the intensity of the light pulse is limited. The latter can improve the intensity of laser pulses effectively. In this paper, an implementation of microwave base on twin-bunch FEL is proposed. We use two basic methods to make sure the beams quality consistency and energy adjustability in linear accelerator.

### The Settable Waveform Table

The LLRF system provides a settable waveform table method to set a step-shaped waveform, which benefits by MTCA.4 system. We set a user defined table with 1024 points data into a BRAM group of FPGA and drive the DAC with this table to generate the arbitrary waveform, a step-shaped waveform is used in the confirmatory experiments. Due to the limited bandwidth of the klystron, the burst signal of the step-shaped pulse may cause the excitation ring, so we set a sigmoid smooth function (as Eq. (2)) between two steps, in case of the power oscillation of the klystron.

$$g(z) = \frac{k}{1+e^{-z}} \quad (2)$$

Because the phase change can be seen as the frequency change, phase saltation may be out of the bandwidth of klystron. Therefore, the waveform must be continuous in the time domain.

### The Pulse Shaping Feed-back Algorithm

The settable waveform also can be modified by a feed-back algorithm to shape the pulse flat (in Fig. 1), is benefited to keep the consistency of two beam energy. Get a new waveform table after 60-80 times iteration (pulse repeat frequency, 20 Hz in DCLS) (in Fig. 2), and save the data into a BRAM group of FPGA. The feed-back algorithm is based on the Eq. (3):

$$\begin{cases} I_{set}(n+1) = P * \frac{I_{ref} - \frac{I_{out}}{sgain}}{kgain} + I_{set}(n) \\ I_{set}(n+1) = P * \frac{I_{ref} - \frac{I_{out}}{sgain}}{kgain} + I_{set}(n) \end{cases} \quad (3)$$

$kgain$ is the slope of klystron working point, $sgain$ is the gain of the power system, $P$ is the proportional parameter. And the architecture of the algorithm is shown in Fig. 1.

The new waveform table with shaping data can combine with the intra-pulse feed-forward algorithm to suppress the jitter. The relational expression follows Eq. (4):

___
* Work supported by the National Natural Science Foundation of China (Grant No. 22288201), the Scientific Instrument Developing Project of the Chinese Academy of Sciences (Grant No. GJJSTD20220001), and the Shenzhen Science and Technology Program (Grant No. RCBS20221008093247072).
† dinghongli@dicp.ac.cn
‡ zhujinfu@mail.iasf.ac.cn

$$\begin{bmatrix} I_{set} \\ Q_{set} \end{bmatrix} = \begin{bmatrix} cos\varphi_n & -sin\varphi_n \\ sin\varphi_n & cos\varphi_n \end{bmatrix} \begin{bmatrix} \frac{1}{kgain} & 0 \\ 0 & \frac{A_n}{A} \end{bmatrix} \begin{bmatrix} -\Delta A \\ -A\Delta\varphi \end{bmatrix} =$$
$$\begin{bmatrix} \frac{-\Delta A cos\varphi_n}{kgain} + Asin\Delta\varphi sin\varphi_n \\ \frac{-\Delta A sin\varphi_n}{kgain} - Asin\Delta\varphi cos\varphi_n \end{bmatrix} \quad (4)$$

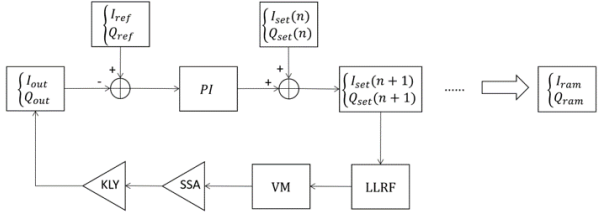

Figure 1: Architecture of the algorithm

The setting waveform and the compensation value both can be seen as a waveform table to control the LLRF output. The amplitude and phase information of step-shaped mode need to be combined into the raw compensation. $A_n$ is the amplitude information in waveform table RAM of the step-shaped mode, and $\varphi_n$ is the phase.

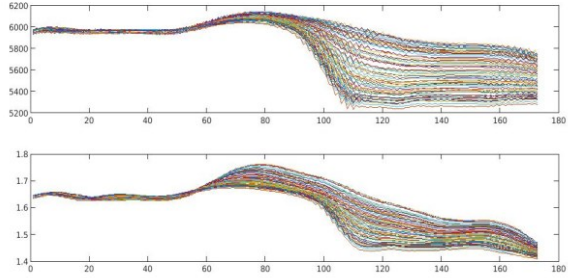

Figure 2: The pulse shaping feed-back operating.

## BEAM EXPERIMENTS

The Tow-colour FEL in pulse mode demands each beam to have the same quality (energy spread, phase, etc.) except the energy and the energy need to be continuously adjustable. Therefore, it has two basic requirements in DCLS device [3] described in Fig. 3.

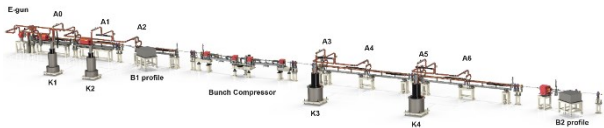

Figure 3: The linear accelerator of DCLS FEL device.

First, each beam needs to keep energy consistency after the E-gun and the cavity A4. For the E-gun system according to the properties of standing wave cavity, the relation of E-gun maximum voltage $V_C$ and forward power $P_f$ (typical 7 MW in DCLS) can be calculated in the equation as below. In addition, a flat-top of $V_C$ need to be set for providing the same energy to each beam.

$$V_C = 2R_L I_f (1 - e^{-t/\tau}) \quad (5)$$

$$I_f = \sqrt{\frac{2\beta}{\beta+1} \cdot \frac{P_f}{R_L}} \quad (6)$$

The $\tau$ is the time constant, $\tau = 1/\omega_{1/2}$, $\beta$ is the coupling factor, $R_L$ is load impedance. According to the Eq. (5) and (6), the E-gun cavity voltage is non-flat when excitation is a square wave as shown in Fig. 4. To set a flap-top of the E-gun cavity voltage, a step-shaped waveform table needs to be established, and the duration of the flat-top is also can be calculated by Eq. (7):

$$t_f = t_w + \tau \cdot \ln\left(1 - \sqrt{\frac{P_{f2}}{P_{f1}}}\right) \quad (7)$$

$P_{f1}$ is the first step power of $P_f$, $P_{f2}$ is the second step power of $P_f$, $t_w$ is the pulse width. It indicates use a step-shaped excitation power can put off the $V_C$ arrival time of the system and generate a flap-top as shown in Fig. 4.

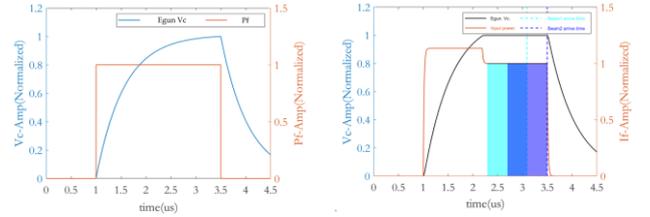

Figure 4: Relation of E-gun voltage and forward power

For DCLS system, the maximum time of flap-top is 1.309us to keep the $V_C$ in typical $P_f$ (improve the first step power to maximum) and the basic parameters of the DCLS can see in Table 1. In consideration of the filling time of cavity and the coupling between E-gun and cavity A0, the gap between two beams is limited in 0 to 350ns for now.

Table 1: Basic parameters of DCLS E-gun system

| Parameter | Value | Unit |
| --- | --- | --- |
| Top | 2.5 | us |
| Bottom | 2856 | MHz |
| E-gun coupling factor | 1($Q_0$=1×$10^4$) | -- |
| Cavity filling time | 800 | ns |
| Klystron maximum power | 9 | MW |

In addition, the pulse shaping feed-back algorithm is used in each klystron system to make the power stay the same intra pulse. To keep the consistency of beams in cavity A1 to A4. But when klystron operate in the saturation power region, the feedback algorithm lose effect on the amplitude.

The second demand is each beam needs to be provided with different energy in the last two accelerated cavities A5 and A6, whose structure is a traveling wave cavity drive by klystron 4. The actual energy of the electron beam in the traveling wave cavity is the integral of the power in cavity filling time [4] (800ns in DCLS). Thanks to the settable waveform table, an arbitrary pulse power mode can be set in the cavity. The step-shaped mode can adjust the amplitude o adjust the energy difference of the beams.

## RESULTS OF THE EXPERIMENTS

Due to the laser device which is used for the pump-probe experiment being in the development process, we set the different delays of the microwave systems to put the beams in different positions, simulate the twin-bunch case.

For E-gun system, settable waveform table is used to generate a flap-top of the E-gun field (see Figs.5 and 6).

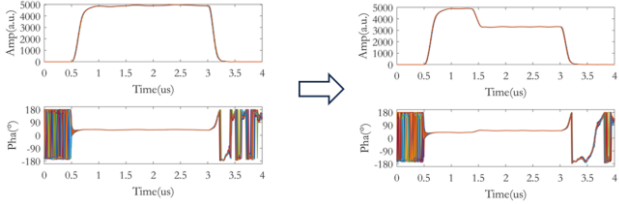

Figure 5: The klystron power amplitude and phase

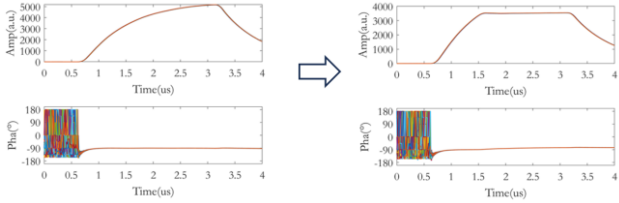

Figure 6: The E-gun voltage amplitude and phase

We set the second step as a lower power in these confirmatory experiments to get a wider flap-flop and protect the klystron. Basically, the step-shaped wave mode can ensure that the energy of two beams can keep the consistency after being ejected by the E-gun ($\Delta t=0\sim350$ns). The figure show the results of the time interval is 100ns (see Fig.7).

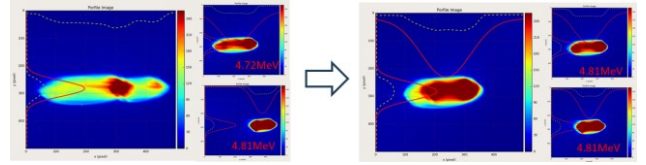

Figure 7: Energy of two beams, $\Delta t=100$ns

And for passing through the accelerated unite drive by the first three klystrons (A0 to A4). The pulse shaping feedback algorithm is used for keeping the beam consistency (presented in Fig. 8). We capture the beam energy by the profile in bunch compressor [5] and prove the ability (see Fig.8). Slight energy spread change caused by the coupling of the phase and amplitude.

To provide two-colour FEL in different wavelengths, the system realizes the tuneable microwave field in the last klystron and changes the electron beams in different energy by setting the step-shaped mode microwave field in klystron4. And detected the two beams power in different position by the last beam-bending magnet profile (shown in Fig.9).

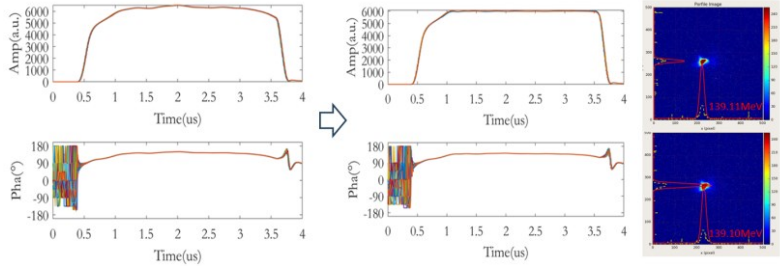

Figure 8: Waveform in K2 and energy of two beams in bunch compressor, $\Delta t=100$ns

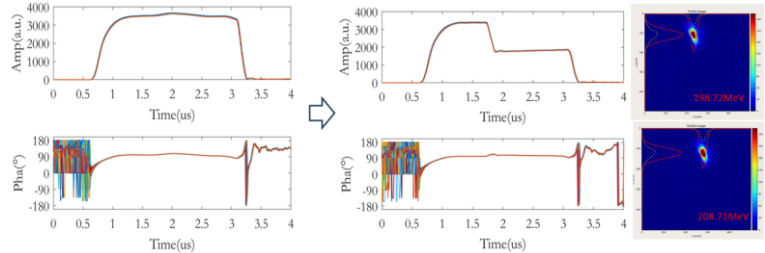

Figure 9: Waveform in K4 and Energy of two beams at the last beam-bending profile, $\Delta t=180$ns

## CONCLUSION

A preliminary design scheme of two-colour FEL microwave is introduced, and a series of beams experiments are carried out based on the DCLS FEL device, which lays a theoretical foundation for the pump-probe experiments execution with two-colour FEL. Combined with the beam experiment, the feasibility of an arbitrary shapes microwave function in DCLS has been proved.

## REFERENCES

[1] S. Serkez *et al.* "Opportunities for two-color experiments in the soft X-ray regime at the European XFEL", in *Appl. Sci. 2020, 10(8), 2728*, Apr. 2020.

[2] A. Marinelli *et al.* "Twin-bunch Two-colour FEL at LCLS", in *Int. Particle Accelerator Conf.(IPAC'16),* Busan, Korea, May 2016, paper TUZA02.

[3] M. Zhang *et al.* "Linac design for dalian coherent light source", in *Proc. 2013 Int. Particle Accelerator Conf. (IPAC2013)*, Shanghai, China, 2013, paper TUPEA042.

[4] M. Paraliev *et al.* "SwissFEL double bunch operation", in *Physical review accelerators and beams 25, 120701 (2022)*, Villigen, Switzerland, 2 Dec. 2022, doi: 10.1103/PhysRevAccelBeams.25.12070

[5] M. Zhang *et al.* "Commissioning experience and beam optimization for DCLS linac", in *Proc. 2017 Int. Particle Accelerator Conf. (IPAC2017)*, Shanghai, China, 2017, paper TUPAB083.